# The structure and properties of graphene supported on gold nanoparticles

*Zoltán Osváth,*[*,§,†] *András Deák,*[§] *Krisztián Kertész,*[§,†] *György Molnár,*[§] *Gábor Vértesy,*[§] *Dániel Zámbó,*[§] *Chanyong Hwang,*[‡,†] *and László P. Biró* [§,†]

[§]Institute of Technical Physics and Materials Science, MFA, Research Centre for Natural Sciences, HAS, 1525 Budapest, P.O. Box 49, Hungary

[†]Korea-Hungary Joint Laboratory for Nanosciences (KHJLN), P.O. Box 49, 1525 Budapest, Hungary

[‡]Center for Nano-metrology, Division of Industrial Metrology, Korea Research Institute of Standards and Science, Yuseong, Daejeon 305-340, Republic of Korea

**Abstract**

Graphene covered metal nanoparticles constitute a novel type of hybrid materials, which provide a unique platform to study plasmonic effects, surface-enhanced Raman scattering (SERS), and metal-graphene interactions at the nanoscale. Such a hybrid material is fabricated by transferring graphene grown by chemical vapor deposition onto closely spaced gold nanoparticles produced on a silica wafer. The morphology and physical properties of nanoparticle-supported graphene is investigated by atomic force microscopy, optical reflectance spectroscopy, scanning tunneling microscopy and spectroscopy (STM/STS), and confocal Raman spectroscopy. This study shows that the graphene Raman peaks are enhanced by a factor which depends on the excitation wavelength, in accordance with the surface plasmon resonance of the gold nanoparticles, and also on the graphene-nanoparticle distance which is tuned by annealing at moderate temperatures. The observed SERS activity is correlated to the nanoscale corrugation of graphene. STM and STS measurements show that the local density of electronic states in graphene is modulated by the underlying gold nanoparticles.

**Introduction**

The exceptional electronic, thermal, and mechanical properties of graphene[1] make this two-dimensional material an ideal platform for building a series of graphene based functional nanomaterials. Hybrid structures made of graphene and metal nanoparticles[2] are a class of nanocomposites which can display novel physical properties by combining the unique properties of graphene and the advantages of metallic nanoparticles.[3] For example, the localized surface plasmon resonance (LSPR) of metallic nanoparticles[4] has been used to enhance the light absorption of graphene and greatly increase photocurrents in graphene-based photodetectors.[5,6] It is known that local electric fields are enhanced at the LSPR, and this is one of the main mechanisms of surface-enhanced Raman spectroscopy (SERS).[7] A number of studies exploit the possibility to integrate graphene with plasmonic gold[8–14] or silver nanoparticles[15–18] in order to elaborate SERS platforms with improved performance. Graphene-covered metal nanoparticles can have favourable characteristics as SERS substrates due to the protective graphene layer, which prevents metal nanoparticles from oxidation.[19,20] Not only the LSPR of metal nanoparticles enhances scattering of graphene, but inversely, graphene can be used to tune the LSPR frequency of metal nanostructures.[21] This can be controlled for example by introducing a spacer layer between graphene and gold nanoparticles (Au NPs).[22] Graphene combined with plasmonic nanoparticles has been demonstrated to provide real potential for applications in catalysis, biosensors, fuel cells,[2] as well as in novel graphene-based optoelectronic devices such as light-emitting diodes, solar cells,[23] or advanced transparent conductors.[24] Therefore, the detailed microscopic characterization of such hybrid materials is of great importance. In this work we study the properties of graphene transferred on top of Au NPs with heights of 15 – 20 nm prepared by evaporation of gold on $SiO_2$ substrate and subsequent annealing. Thorough characterization was performed using atomic force microscopy (AFM), confocal Raman and optical reflectance spectroscopy, as well as scanning tunneling microscopy and spectroscopy (STM/STS). We correlate the observed SERS activity to the nanoscale corrugation of graphene and provide new insights into the optical and local electronic properties of gold nanoparticle-supported graphene.

**Experimental**

Gold nanoparticles were prepared as follows: gold grains of 99.99% purity were applied as source material for evaporation, which was carried out from an electrically heated tungsten boat, at a background pressure of $5\times10^{-7}$ mbar. A thin gold film of 5 nm was evaporated onto a 285-nm-$SiO_2$/Si substrate at a rate of 0.1 nm/s. During evaporation the substrate was held at room temperature. Subsequent annealing was performed at 400 °C in Ar atmosphere for 30 minutes, which resulted in the formation of Au nanoparticles with heights of 15-20 nm and high surface coverage.

Large-area graphene was grown on a mechanically and electro-polished copper foil (25 μm thick, 99.8% purity, Alfa-Aesar), which was inserted into a thermal CVD furnace. The furnace was evacuated to ~$10^{-4}$ mbar and the temperature was raised to 1010 °C with $H_2$ gas flow (~$10^{-2}$ mbar). When the temperature became stable, both $CH_4$ (20 sccm) and $H_2$ (5 sccm) were injected into the furnace for 8 minutes to synthesize the graphene. After the growth, we cooled down the furnace with a cooling rate of 50 °C/min. The graphene sample was transferred onto the Au NPs using thermal release tape, and an etchant mixture consisting of $CuCl_2$ aqueous solution (20%) and hydrochloric acid (37%) in 4:1 volume ratio. After the etching procedure, the tape holding the graphene was rinsed in distilled water, then dried and pressed onto the surface covered by nanoparticles. The tape/graphene/Au NPs/$SiO_2$/Si sample stack was placed on a hot plate and heated to 95 °C, slightly above the release temperature of the tape. The tape was removed, leaving behind the graphene on top of Au NPs.

The graphene/Au NPs hybrid structure was investigated both before and after annealing by confocal Raman microscopy (WITec) using excitation lasers of 488 and 633 nm. Reference spectra from graphene/$SiO_2$ were recorded on areas not covered by Au NPs of the same sample. Atomic force microscopy was performed in a MultiMode 8 AFM (Bruker) operating in tapping mode under ambient conditions.

A similar sample was prepared on HOPG substrate for scanning tunneling microscopy. First, the HOPG was irradiated at normal incidence in an ion implanter with $Ar^+$ ions of 60 keV using a dose of D = 8.4 × $10^{15}$ $cm^{-2}$. The irradiation was applied in order to modify the atomically flat surface of the HOPG and to make its roughness comparable to the surface roughness of the $SiO_2$. Then, we deposited 5 nm of Au and applied subsequent annealing at 400 °C in Ar atmosphere for 30 minutes. We obtained Au NPs which are similar in both size and surface coverage to the ones prepared on $SiO_2$/Si substrate. Finally, graphene was transferred onto the Au NPs using the method described above. STM and STS measurements were performed in a DI Nanoscope E operating under ambient conditions.

Optical reflectance spectra were taken using Avantes 1024X122 TEC fibre optic spectrometer. We used a pair of fibre probes for illumination and detection, each with 200 μm in diameter. We measured the reflectance signal at different reflected angles, under constant perpendicular illumination (Avantes DH balanced light source). White Avantes reference sample was used.

**Results and Discussion**

We used tapping mode AFM to investigate the size and shape of the prepared gold nanoparticles, as well as to image the graphene transferred onto these. The surface of a thin gold film of 5 nm deposited on $SiO_2$ is shown in Fig. 1a. The surface transforms by annealing into a film of closely spaced gold nanoparticles (Fig. 1b) due to the diffusion and aggregation of gold clusters.[25]

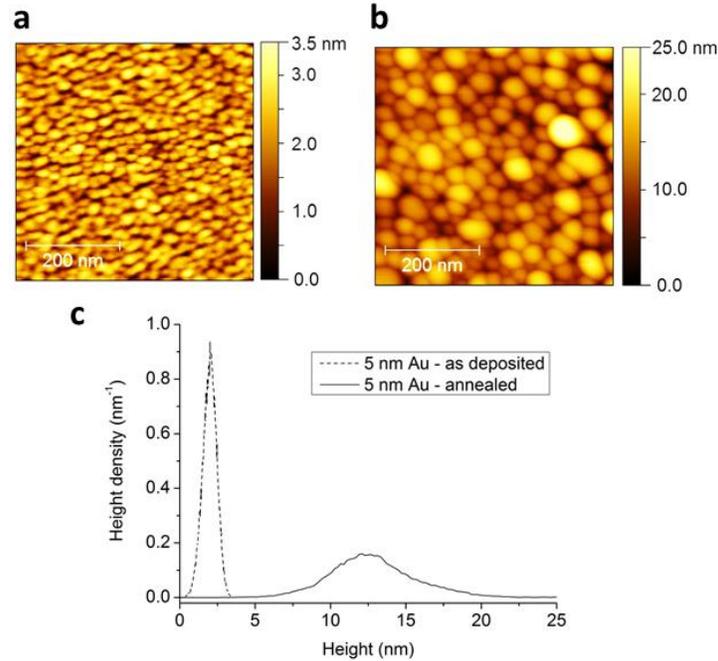

***Figure 1.*** *Tapping mode AFM image of Au(5 nm)/SiO$_2$ (a) as deposited and (b) after annealing at 400 $^o$C. (c) Height distributions corresponding to (a) – dashed line, and (b) – solid line, respectively.*

During this process the root mean square (RMS) roughness increases from 0.5 nm to 3 nm. The corresponding height distributions are displayed in Fig. 1c. The majority of Au NPs have heights between 9 and 16 nm, while the average height is 12.5 nm (solid line in Fig. 1c). We transferred graphene grown by chemical vapour deposition (CVD) onto the prepared gold nanoparticles as described in the Experimental section. Figure 2a shows a typical AFM image of the transferred graphene which is considerably rippled. Note, that the lower part of the image is not covered with graphene.

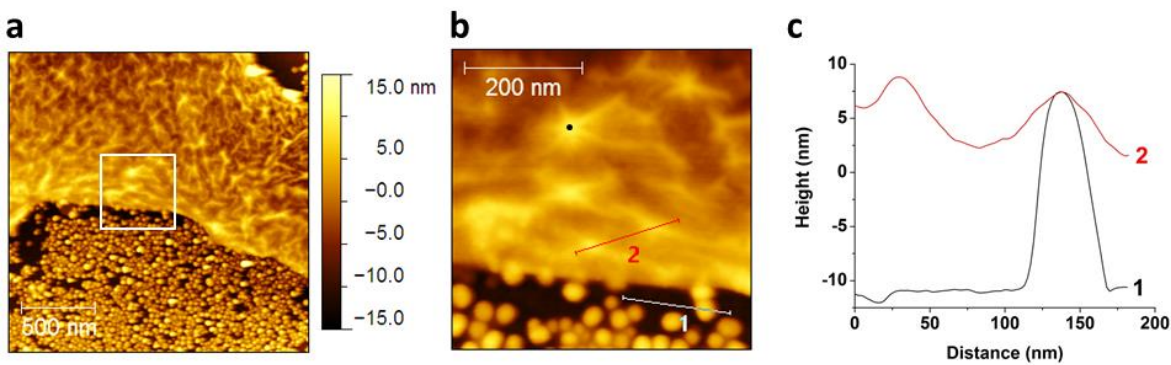

*Figure 2.* Tapping mode AFM image of graphene transferred onto Au nanoparticles. The area marked by a white square in (a) is presented with higher magnification in (b). The black dot in (b) points out star-shaped rippling centred on the top of the underlying nanoparticle. The height profiles corresponding to the line section 1 and 2 are displayed in (c).

During the transfer with thermal release tape, the initial large area graphene breaks into smaller sheets with dimensions of several micrometers and not all of them remain attached to the nanoparticles. Figure 2b is a higher magnification image which corresponds to the white square drawn in Fig. 2a. The height profiles corresponding to the line section 1 and 2 are displayed in Fig. 2c. Line section no. 1 is measured in the area without graphene, showing a typical gold nanoparticle on the $SiO_2$ surface, with height of 18 nm. On the graphene-covered side, the line section no. 2 displays the wavy shape of the graphene. The peaks in the height profile correspond to graphene directly supported by nanoparticles, whereas the dip corresponds to graphene bridging two nanoparticles. Comparing the height profiles of line section 1 and 2, we find that the graphene part bridging the nanoparticles is located more than 10 nm above the $SiO_2$ substrate, i.e. it is suspended. In fact, this is a general observation for the transferred graphene: it is either supported by gold nanoparticles, either suspended between them, but never touching the underlying substrate. This is similar to the structure of graphene transferred onto $SiO_2$ nanoparticles with comparable diameter.[26] Ripples developed in the shape of a star are often observed (see Fig. S1 in the Electronic Supplementary Information (ESI)). Such star-shaped ripples, like the one marked in Fig. 2b (black dot), develop around a supporting nanoparticle, which exerts a mechanical force on the graphene membrane.

AFM measurements show that annealing at moderate temperatures affects significantly the nanoscale corrugation of nanoparticle-supported graphene. This is demonstrated in Fig. 3, where the AFM image of the same graphene area is displayed both without annealing (Fig. 3a), and after annealing at 500 °C in $N_2$ atmosphere (Fig. 3b).

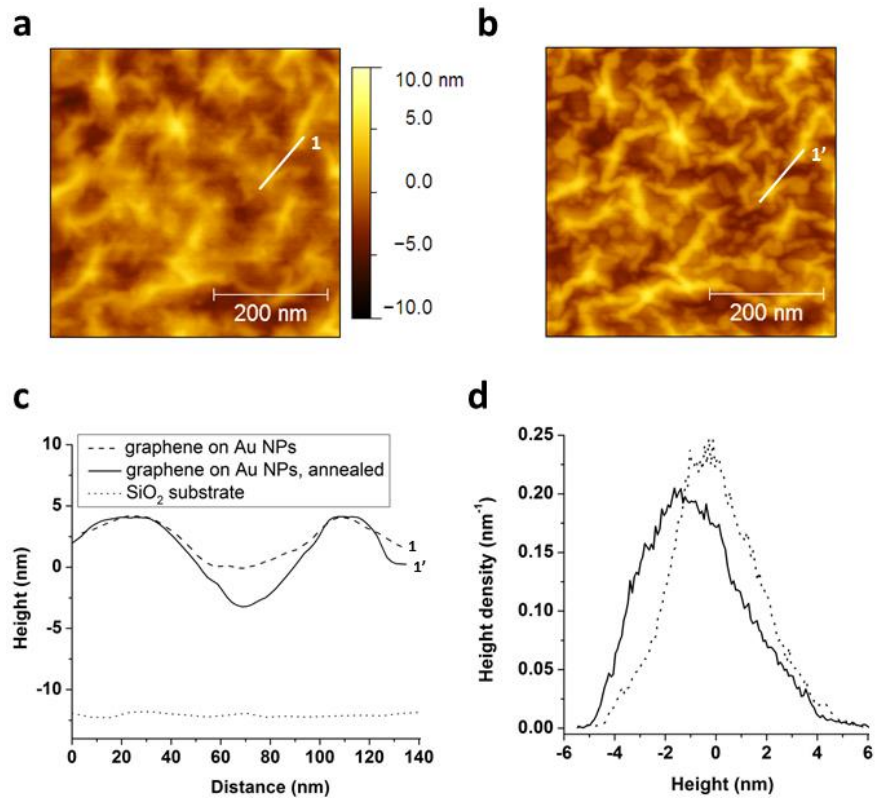

*Figure 3.* *Effect of annealing at moderate temperature. AFM images measured on the same area of nanoparticle-supported graphene (a) before and (b) after annealing at 500 °C. (c) Height profiles 1 and 1' taken along the same line section marked with white line in (a) and (b), before (dashed line) and after annealing (solid line), respectively. As a reference, the $SiO_2$ substrate level is also shown (dotted line). (d) Height distributions corresponding to a) (dotted line), and b) (solid line). The height at 0 nm in c)-d) corresponds to the average height of a).*

Observe that the surface texture of the graphene is more pronounced in Fig. 3b. The height profiles 1 and 1' in Fig. 3c are taken along the same line section marked in Fig. 3a-b (white line). The vertical distance between the saddle points (suspended graphene regions) of the height profile 1 and 1' is 3.1 nm. This shows that graphene is more rippled after annealing, i.e. the suspended regions penetrate more deeply into the space between the nanoparticles. As a reference, the height level of the bare $SiO_2$ surface – extracted from a lower magnification AFM image – is also plotted in Fig. 3c (dotted line). Considering the whole area shown in Fig. 3, we find that after annealing the graphene membrane is closer to gold nanoparticles by 0.8 nm, on average. This is demonstrated by the shift of the corresponding height distributions shown in Fig. 3d.

It has been shown recently[27] that the distance between graphene and an underlying thin gold film determines the surface-enhanced Raman scattering (SERS) properties of the graphene/gold substrate. We investigated the SERS activity of the graphene/gold nanoparticle sample by confocal Raman spectroscopy performed both before and after annealing. Figure 4a show typical Raman spectra obtained with 488 nm laser on transferred CVD-grown graphene without annealing. The 2D peak at $2628\ cm^{-1}$ has a full width at half maximum of $28\ cm^{-1}$, which corresponds to a single component peak, characteristic of monolayer graphene. Note that there is no significant difference between graphene peak intensities when measured on $SiO_2$ and on gold nanoparticles, respectively. In contrast, when using the 633 nm laser we observe an almost tenfold enhancement (Fig. 4b) for the graphene G peak ($1580\ cm^{-1}$), as well as 4-fold enhancement for the 2D peak. After annealing, these enhancement factors increase to 13 and 22 for the G and the 2D peak, respectively (Fig. 4d). A significant increase of the D-peak ($1318\ cm^{-1}$) is also observed. Furthermore, nearly 6-fold peak enhancement is observed with the 488 nm laser also, but only for the 2D peak (Fig. 4c). Analysing the spectra measured after annealing, we find that the G and 2D peak positions corresponding to graphene on Au NPs are downshifted compared to the peak positions obtained from graphene on $SiO_2$ (for details see Fig. S2 in the ESI). This is in agreement with recent measurements obtained from graphene placed on top of two closely spaced gold nanodisks,[28] and it is attributed to tensile strain induced in graphene by the underlying Au NPs.

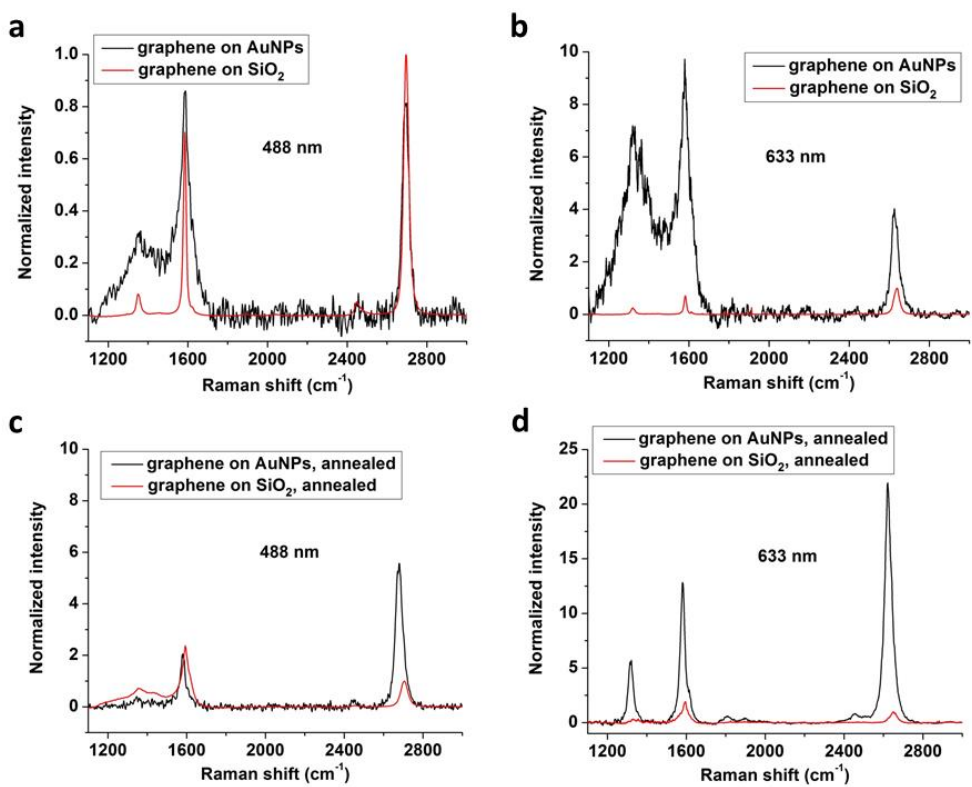

*Figure 4.* Graphene Raman spectra averaged over areas of 5×5 µm². The spectra denoted with black line correspond to graphene transferred on gold nanoparticles, while the spectra in red are acquired on graphene transferred directly onto $SiO_2$ substrate. The measurements were performed a)-b) before and c)-d) after annealing, using excitation lasers of 488 and 633 nm, respectively. The fluorescent background from gold nanoparticles was removed in all cases. All spectra are normalized to the 2D peak height measured on $SiO_2$.

To better understand the SERS activity of the graphene/Au NPs, we now focus on the optical reflectance properties depicted by the relative reflectance spectra shown in Fig. 5a, which were recorded in perpendicular illumination and detection configuration. Considering the reflectance of the Au NPs on $SiO_2$ substrate, without graphene (dotted line), the spectral minimum at 597 nm corresponds to the LSPR of gold nanoparticles, where light is highly absorbed. For reference, the spectrum of graphene/$SiO_2$ is also displayed (dashed line).

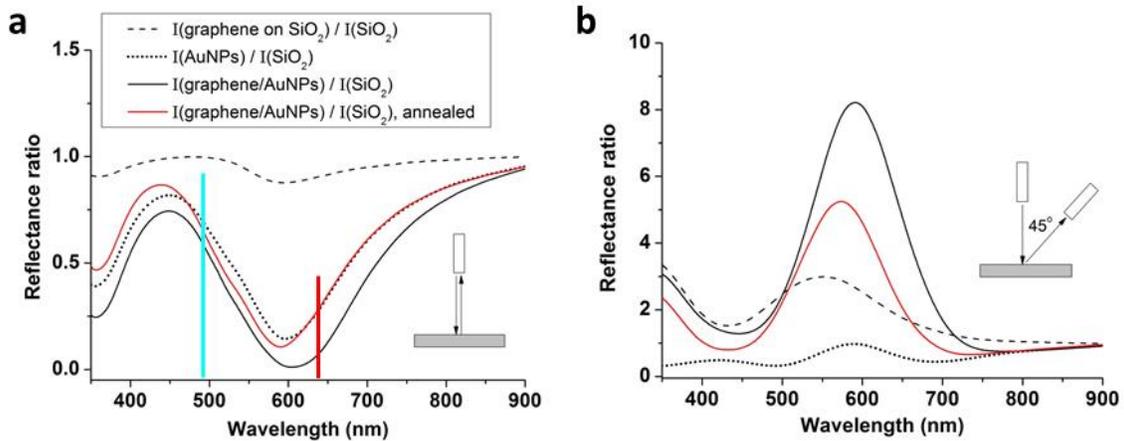

*Figure 5.* *Optical reflectance spectra of graphene samples. All spectra are divided by the spectrum recorded from bare $SiO_2$ surface. The spectra of graphene/Au NPs are shown before and after annealing, with black and red solid lines, respectively. For reference, the spectra of graphene/$SiO_2$ and Au NPs/$SiO_2$ are also displayed with dashed and dotted lines, respectively. a) Reflectance spectra measured under perpendicular illumination and detection. The coloured vertical lines correspond to the laser wavelengths used for Raman spectroscopy (488 and 633 nm). b) Reflectance measurements under perpendicular illumination and detection angle of $45^o$.*

When graphene is transferred on the Au NPs, the LSPR peak broadens and red-shifts about 9 nm (black solid line). This is in agreement with observations performed on graphene transferred onto silver nanoparticles,[18] and it is due to the coupling between localized surface plasmons and graphene. As recent calculations based on dipole approximation show,[22] antiparallel image dipole is formed in graphene when the distance between Au NPs and graphene is small. This image dipole reduces the internal field in the Au NPs, which results in a red shift of the resonance wavelength. Note that the 633 nm laser (red vertical line in Fig. 5a) is closer to the resonance wavelength of the graphene/Au NPs. Consequently, when this laser is used for Raman spectroscopy, it yields stronger SERS activity both before and after annealing, as compared to 488 nm laser (cyan vertical line).

In addition to the observed red shift in the LSPR, there is also a general decrease of the reflectance at all wavelengths compared to the reflectance from Au NPs on $SiO_2$. Similar decrease of reflectance was observed recently on graphene-covered gold nanovoid arrays[29] and it was attributed to enhanced light absorption of graphene. Nevertheless, we find that significant light is scattered also in lateral direction, which actually decreases the intensity of perpendicular reflectance. Fig. 5b shows the reflectance spectra corresponding to perpendicular illumination and detection angle of 45° (for intermediate detection angles see Fig. S3 in the ESI). Near the LSPR, the intensity of the scattered light from graphene/Au NPs (black solid line) is about 8 times the intensity of bare Au NPs (dotted line). This is attributed to the rippled structure of graphene. Since the graphene plane is not horizontal, the generated image dipoles can re-emit the absorbed light under different angles. Moreover, the wave-guiding ability of graphene[30,31] can also influence the scattering process. The light which is coupled to propagating modes supported by graphene is scattered out from a different point of the sample. Note that after annealing, this lateral scattering from graphene/Au NPs is reduced (red spectrum in Fig. 5b), and consequently the perpendicularly detected reflectance is increased (red spectrum in Fig. 5a). Interestingly, the minimum of the LSPR (590 nm) is blue-shifted about 16 nm compared to the minimum measured from graphene/Au NPs before annealing (black solid line). We recall that the average distance between graphene and Au NPs decreased after annealing, as shown by AFM measurements. However, such decreased separation should result in a red shift of the resonance wavelength due to increased plasmonic coupling with graphene.[22] Nevertheless, the decreased separation affects also the electrostatic doping of graphene from the Au NPs. Both the amount and the sign of charge transfer depend on the graphene-gold equilibrium distance.[32] It was demonstrated recently[33] that the near infrared plasmon resonance in a graphene-gold nanorod system can be tuned by controlling the doping in graphene. We thus infer that the observed blue shift in the LSPR should be related to the annealing-induced changes in the doping level of graphene. Additionally, the change in the nanoscale corrugation of graphene observed by AFM can also affect the scattering properties of graphene/Au NPs.[34] The angle between the electric field of LSPR and suspended graphene regions increases upon annealing, which should modify the polarization of graphene by the same LSPR field. Additionally, as graphene sticks better, and fills more space between the nanoparticles after annealing, the effective refractive index of the graphene/Au NP system should change as well. This requires further investigations and goes beyond the scope of the current work.

Next, the LSPR of the gold nanoparticles prepared on $SiO_2$ substrate is studied in more details. Figure 6a shows a typical AFM image of several Au NPs. We performed height profile analysis on this AFM image to extract the nanoparticle diameters and heights, as well as the distances between nanoparticles. Based on the extracted parameters we reconstructed the measured group of nanoparticles as closed dome structures using MATLAB® (Fig. 6b). The extinction spectrum of the reconstructed Au NPs was simulated using boundary element method (BEM),[35] neglecting substrate effects. The simulated extinction spectrum yields an LSPR maximum at 564 nm (Fig. 6c), which is in very good agreement with the LSPR maximum obtained from the extinction measurement (560 nm) performed in a transmission geometry on Au NPs prepared on glass substrate by the same method.

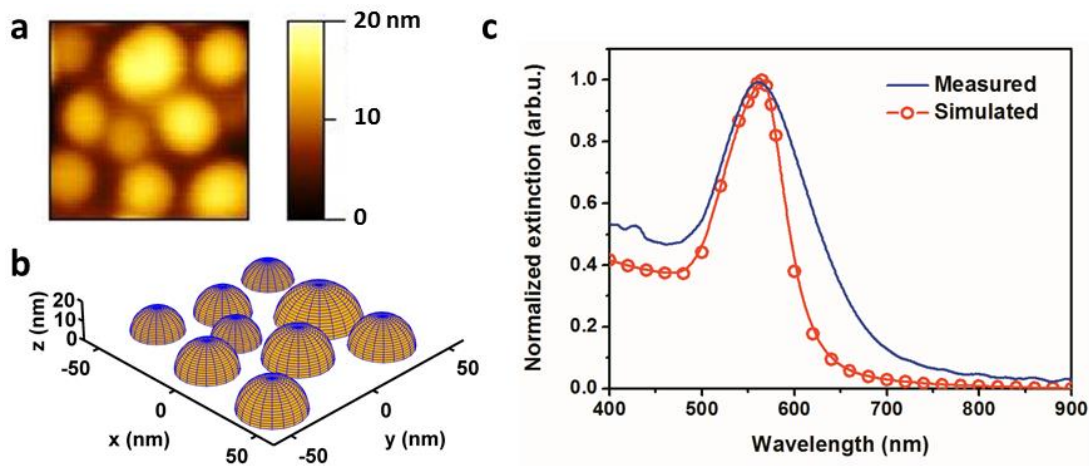

*Figure 6.* a) AFM image (100×100 nm$^2$) of gold nanoparticles formed on $SiO_2$ surface. b) Reconstruction of the Au NPs based on the AFM image in a) using MATLAB®. c) Normalized extinction simulated for the group of 9 nanoparticles shown in b) using boundary element method. Extinction measurements (blue solid line) were performed on Au NPs prepared by the same method on glass substrate.

The simulated LSPR is narrower than the measured one due to the limited number of simulated particles and geometrical arrangements. Note that for Au NPs prepared on $SiO_2$ substrate, the measured LSPR maximum (reflectance minimum) is obtained at 597 nm (Fig. 5). This agrees well with the 590 nm resonance wavelength simulated using finite element method (FEM) implemented in COMSOL® for a single particle by taking the substrate also into account (see Fig. S4 in the ESI). Considering that the Au NPs are not spherical, as considered recently,[11] but have dome-like morphology, the influence of graphene on the extinction spectra cannot be taken into account simply by using an analytical expression in the BEM calculations for the effective polarizability of nanoparticles. Nevertheless, the BEM calculations provide a fast and convenient approach to grasp the fundamental difference in the near-field distribution around the Au NPs at the different excitation wavelengths used in the Raman experiments. In Fig. 7 the electric field enhancement $|E|^2/|E_0|^2$ is shown for the reconstructed cluster at different heights above the substrate level (12 nm and 16 nm, respectively), for the two excitation wavelengths used in the Raman experiments (488 nm and 633 nm, respectively).

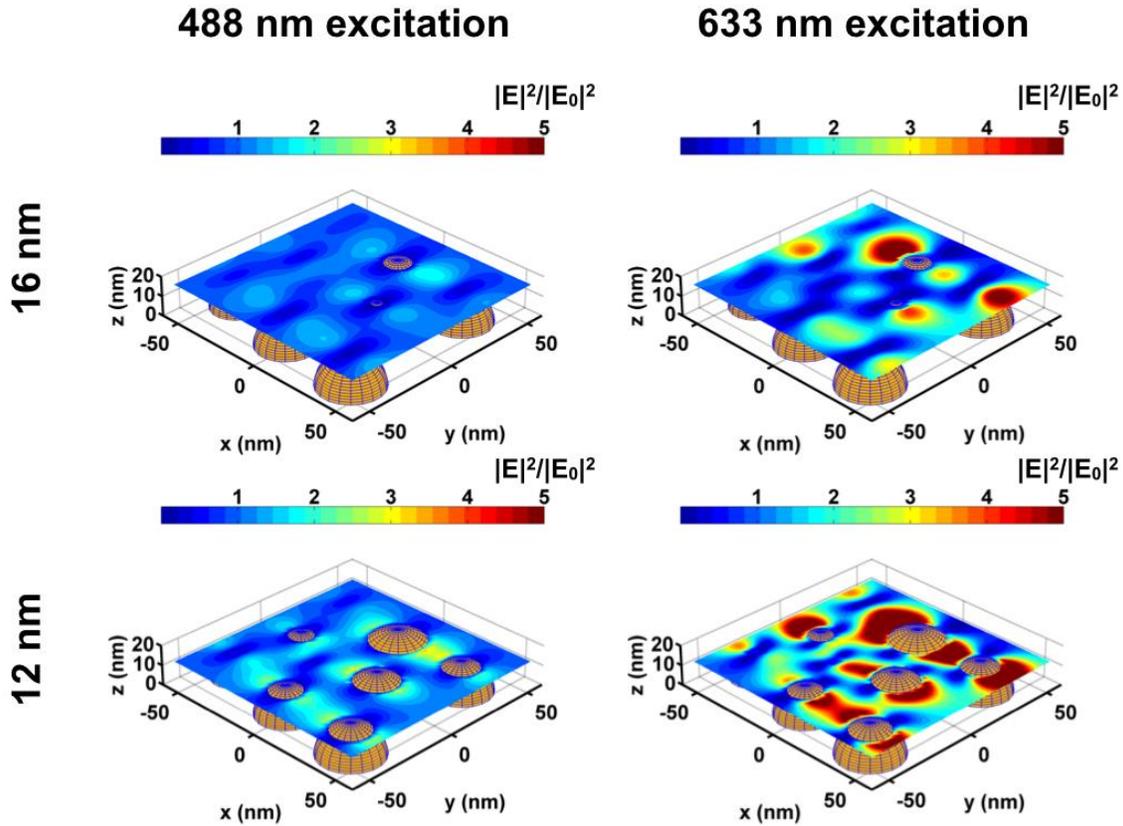

*Figure 7. Near-field distribution around the reconstructed particles from Figure 6b. The intensity maps are shown for the two excitation wavelengths (488 nm and 633 nm, respectively) at two different heights (12 nm and 16 nm) above the substrate level. Light polarization is parallel to the x-axis. The same intensity range is used for all plots. For 633 nm excitation, the field enhancement well exceeds the value of 5 in some regions between two neighbouring particles (hot spots).*

It can be inferred that independently of the excitation wavelength, the deposited graphene monolayer should generally experience a larger electric field as it penetrates deeper into the voids between the nanoparticles due to annealing. For the wavelength closer to the LSPR resonance (633 nm – right panel in Fig. 7) the local increase of the electric field is much more pronounced compared to the off-resonance (488 nm – left panel in Fig. 7) excitation. It is also obvious that the high density of nanoparticles is beneficial for enhancing the Raman signal of graphene. The volume with enhanced electric field extends well into the space between neighbouring particles, especially if the excitation wavelength is closer to the resonance.

Finally, the structure and local electronic properties of graphene/Au NPs was investigated by STM and STS. We prepared similar Au NPs on irradiated highly oriented pyrolytic graphite (HOPG), which served as a conductive surface (see Experimental section). An STM image of graphene transferred on top of Au NPs is shown in Fig. 8a.

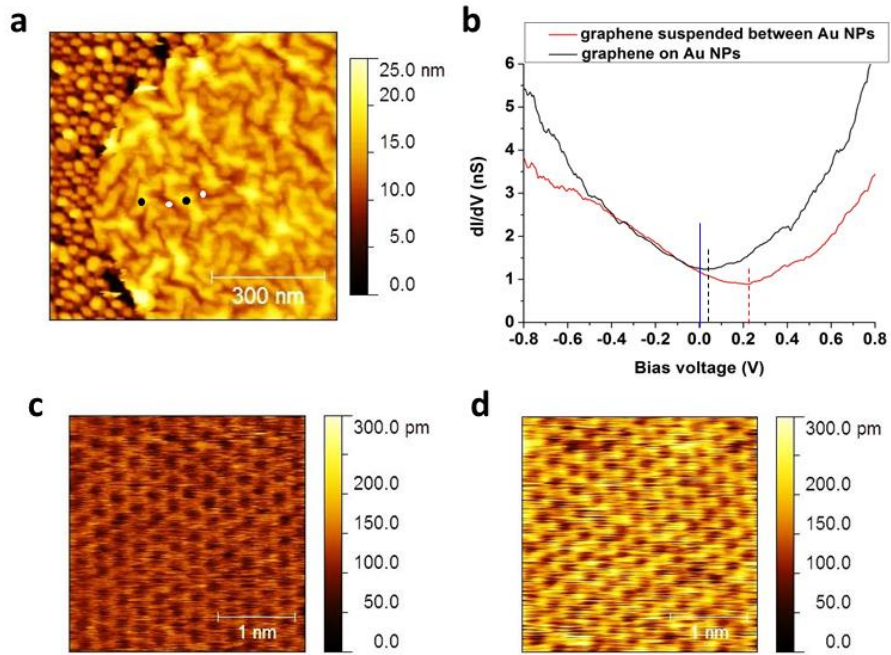

*Figure 8.* a) STM image of graphene/Au NPs. The left part shows uncovered Au NPs. Nanoparticle-supported and suspended graphene regions are marked with black and white dots, respectively. b) Average dI/dV spectra obtained from nanoparticle-supported (black) and suspended (red) graphene regions. The corresponding Dirac points, as well as the Fermi energy (0 V) are marked with vertical lines. As measured atomic resolution STM images of c) nanoparticle-supported and d) suspended graphene. Tunneling parameters: $V_{bias}$ = 20 mV, I = 2 nA.

The structure is similar to the one presented in Fig. 2. One can easily identify the rippled graphene membrane and its edge. Uncovered Au NPs can be observed on the left part of the image. STS measurements were performed on graphene directly supported by Au NPs (black dots), and also on suspended graphene bridging the nanoparticles (white dots). The corresponding average *dI/dV* spectra are shown in Fig. 8b. The Dirac point of suspended graphene (red, dashed vertical line) is 190 mV shifted compared to the Dirac point of supported graphene (black, dashed vertical line). This shows significant difference in the electrostatic doping between the two regions. In supported graphene, the Fermi energy is 30 meV below the Dirac point, and the small p-doping is dominated by the underlying Au NPs. In contrast, the suspended regions are 20 – 30 nm away from the gold nanoparticles, and thus the measured spectra are determined by environmental p'-doping. We infer that the Au NPs have little or no effect on the local density of states (LDOS) of suspended regions. Furthermore, since environmental doping affects the LDOS of both regions, the observed difference between the two Dirac points is attributed to charge transfer from Au NPs to supported graphene regions.[36] Thus, graphene is selectively doped electrostatically, forming a mesh of nanoscale p-p' junctions.

Atomic resolution STM images obtained on these regions (Fig. 8c) show a honeycomb lattice, which is typical for monolayer graphene. We further observed that the corrugation of the atomic resolution images performed in the two regions is different, as shown in Fig. 8c-d (same colour scale). Note that suspended graphene is more difficult to investigate by STM,[37] and thus the atomic resolution image in Fig. 8d is noisier. Nonetheless, typical peak-to-peak distances in the vertical tip movement are around 50 pm in the case of nanoparticle-supported graphene (Fig. 8c), while this value is around 100 pm for suspended graphene areas (Fig. 8d). Here, peak-to-peak distances due to increased noise are not taken into account. The larger corrugation is likely due to tip-sample interaction.[38] The attractive force experienced by graphene towards the STM tip is position-sensitive, distinguishing carbon sites from the centres of hexagonal rings. This force pulls up the carbon atoms, increasing the tunnelling current. Since suspended graphene is more easily pulled than supported graphene, it results larger corrugation in the STM images. The LDOS near the Fermi energy, corresponding to the two regions can be estimated by:[39,40]

$$N_i(E_{F_i}) = 2|E_{F_i}|/(\pi \hbar^2 v_F^2),  \quad (1)$$

where $\hbar$ is the reduced Planck constant, $v_F$ the Fermi velocity, $E_{F_i} = eV_{D_i}$ the Fermi energies measured from the corresponding Dirac points, and $i = 1,2$ is the index for supported and suspended graphene, respectively. Using $V_{D_1} = 30$ mV, and $V_{D_2} = 220$ mV, eqn (1) yields $N_2 = 7.3 \times N_1$, provided that $v_F$ is the same in the two regions. This higher LDOS is somewhat hidden in Fig. 8b, because the minimum tunnelling conductance of suspended graphene is lower than the minimum conductance measured on Au NP-supported graphene.

**Conclusions**

In conclusion, AFM and STM measurements reveal the nanoscale structure of gold nanoparticle-supported graphene, which can be modified by annealing at moderate temperatures. Graphene is completely separated from the SiO$_2$ or HOPG substrates. It is either directly supported by nanoparticles, either suspended between Au NPs. In spite of the relatively small nanoparticle dimensions, the studied graphene/Au NPs material is SERS active, with a maximum enhancement factor of 22 for the graphene 2D peak. The observed SERS effect depends on the laser excitation wavelength and it is attributed to the near-field enhancement around plasmonic Au NPs, as demonstrated by simulations using boundary element method on a group of 9 nanoparticles reconstructed from AFM measurements. We found that the Au NPs have dome-like morphology, which we used also for the calculation of the extinction spectrum. The simulated LSPR maximum is in very good agreement with the measured one.

STS measurements reveal that the LDOS of graphene depends on the spatial position. Suspended graphene regions are more p-doped than supported ones. Thus, graphene is selectively doped electrostatically and forms a network of p-p' nanojunctions. Finally, optical reflectance spectra show that the presence of graphene increases significantly the lateral scattering of the graphene/Au NPs, while the reflectance can be tuned by annealing. We suggest that besides doping effects, the nanoscale corrugation of graphene also affects the reflectance properties of graphene/Au NPs. This is highly intriguing for further theoretical and experimental investigations, since it can open a route towards tailoring the optical properties of graphene/plasmonic nanoparticle hybrid structures through their morphology.

AUTHOR INFORMATION

**Corresponding Author**

*E-mail: osvath@mfa.kfki.hu (Zoltán Osváth)

**Acknowledgment**

The research leading to these results has received funding from the Korea-Hungary Joint Laboratory for Nanosciences and the People Programme (Marie Curie Actions) of the European Union's Seventh Framework Programme under REA grant agreement n° 334377. The OTKA grants K-101599, PD-105173 in Hungary, as well as the János Bolyai Research Fellowships from the Hungarian Academy of Sciences are acknowledged.

**Notes and references**

† Electronic Supplementary Information (ESI) available: [AFM images showing star-shaped graphene rippling, detailed analysis of Raman spectra, additional optical reflectance spectra, and extinction cross section simulations using finite element method]. See DOI: 10.1039/C5NR00268K.